\documentclass[aps, prb, reprint, twocolumn, superscriptaddress]{revtex4-2}

\usepackage{graphicx}
\usepackage{bm}
\usepackage{amsmath}
\usepackage{amssymb}
\usepackage{xcolor}

\usepackage{xcolor}
\usepackage[]{changes}

\definechangesauthor[color=blue]{common}
\definechangesauthor[color=red, name={Paulo Santos}]{PVS}
\definechangesauthor[color=violet, name={Georgy Astakhov}]{GA}
\definechangesauthor[color=purple, name={Alberto Hernandez-Minguez}]{AHM}

\newcommand{\um}[0]{\mu\mathrm{m}}

\newcommand{\wSAW}[0]{\omega_\mathrm{SAW}}
\newcommand{\lSAW}[0]{\lambda_\mathrm{SAW}}

\newcommand{\vBext}[0]{\mathbf{B}}
\newcommand{\Bext}[0]{B}

\newcommand{\Dm}[0]{\Delta m_S}
\newcommand{\mS}[0]{m_S}

\newcommand{\VSi}[0]{\mathrm{V_{Si}}}

\newcommand{\IPL}[0]{I_\mathrm{PL}}

\begin{document}

\bibliographystyle{naturemag}

\title{Acoustically induced coherent spin trapping}
\author{A.~Hern\'{a}ndez-M\'{i}nguez}
\email{alberto.h.minguez@pdi-berlin.de}
\affiliation{Paul-Drude-Institut f\"{u}r Festk\"{o}rperelektronik, Leibniz-Institut im Forschungsverbund Berlin e.V., Hausvogteiplatz 5-7, 10117 Berlin, Germany}
\author{A.~V.~Poshakinskiy}
\affiliation{Ioffe Physical-Technical Institute, Russian Academy of Sciences, 194021 St.~Petersburg, Russia}
\author{M.~Hollenbach}
\affiliation{Helmholtz-Zentrum Dresden-Rossendorf, Institute of Ion Beam Physics and Materials Research, Bautzner Landstrasse 400, 01328 Dresden, Germany}
\affiliation{Technische Universit\"{a}t Dresden, 01062 Dresden, Germany}
\author{P.~V.~Santos}
\affiliation{Paul-Drude-Institut f\"{u}r Festk\"{o}rperelektronik, Leibniz-Institut im Forschungsverbund Berlin e.V., Hausvogteiplatz 5-7, 10117 Berlin, Germany}
\author{G.~V.~Astakhov}
\affiliation{Helmholtz-Zentrum Dresden-Rossendorf, Institute of Ion Beam Physics and Materials Research, Bautzner Landstrasse 400, 01328 Dresden, Germany}

\date{\today}

\maketitle

\textbf{
Hybrid spin-optomechanical quantum systems offer high flexibility, integrability and applicability for quantum science and technology \cite{Arcizet:2011cg, Kolkowitz:2012iw, Barfuss:2015hv,  Schuetz:2015dx, MacQuarrie:2017hl, Lemonde:2018jx, Machielse:2019bt, Poshakinskiy:2019bi}. Particularly, on-chip surface acoustic waves (SAWs) \cite{Delsing_2019} can efficiently drive spin transitions in the ground states (GSs) of atomic-scale, color centre qubits \cite{Golter:2016kf}, which are forbidden in case of the more frequently used electromagnetic fields \cite{Whiteley:2019eu, Maity:2020cn, HernandezMinguez:2020kv}. Here, we demonstrate that strain-induced spin interactions within their optically excited state (ES) can exceed by two orders of magnitude the ones within the GS. This gives rise to novel physical phenomena, such as the acoustically induced coherent spin trapping (CST) unvealed here. The CST manifests itself as the spin preservation along one particular direction under the coherent drive of the GS and ES by the same acoustic field. Our findings provide new opportunities for the coherent control of spin qubits with dynamically generated strain fields that can lead towards the realization of future spin-acoustic quantum devices \cite{Schuetz:2015dx, Lee:2017gs, Wang:2020fh}.  
}

Spin levels in the optically accessible ESs of color centres give rise to their polaronic character \cite{Udvarhelyi:2020fq}, giant thermal shift \cite{Anisimov:2016er} and Landau-Zener-St\"uckelberg interference \cite{Lukin:2020jb}. However, the strong interaction of the ESs with acoustic fields remains largely unexploited. A remarkable example is the CST, which is theoretically proposed and experimentally realized here. It  bears conceptual analogies to electromagnetically induced transparency. 
If the acoustic driving field is tuned in such a way that the spin precesses around the same axis in the GS and ES, the optically detected spin resonance quenches due to the spin trapping along the precession axis. A clear experimental indication of the CST is a Fano-like shape \cite{Limonov2017} of the optically detected spin-acoustic resonance (SAR).


\begin{figure}
\includegraphics[width=0.99\linewidth]{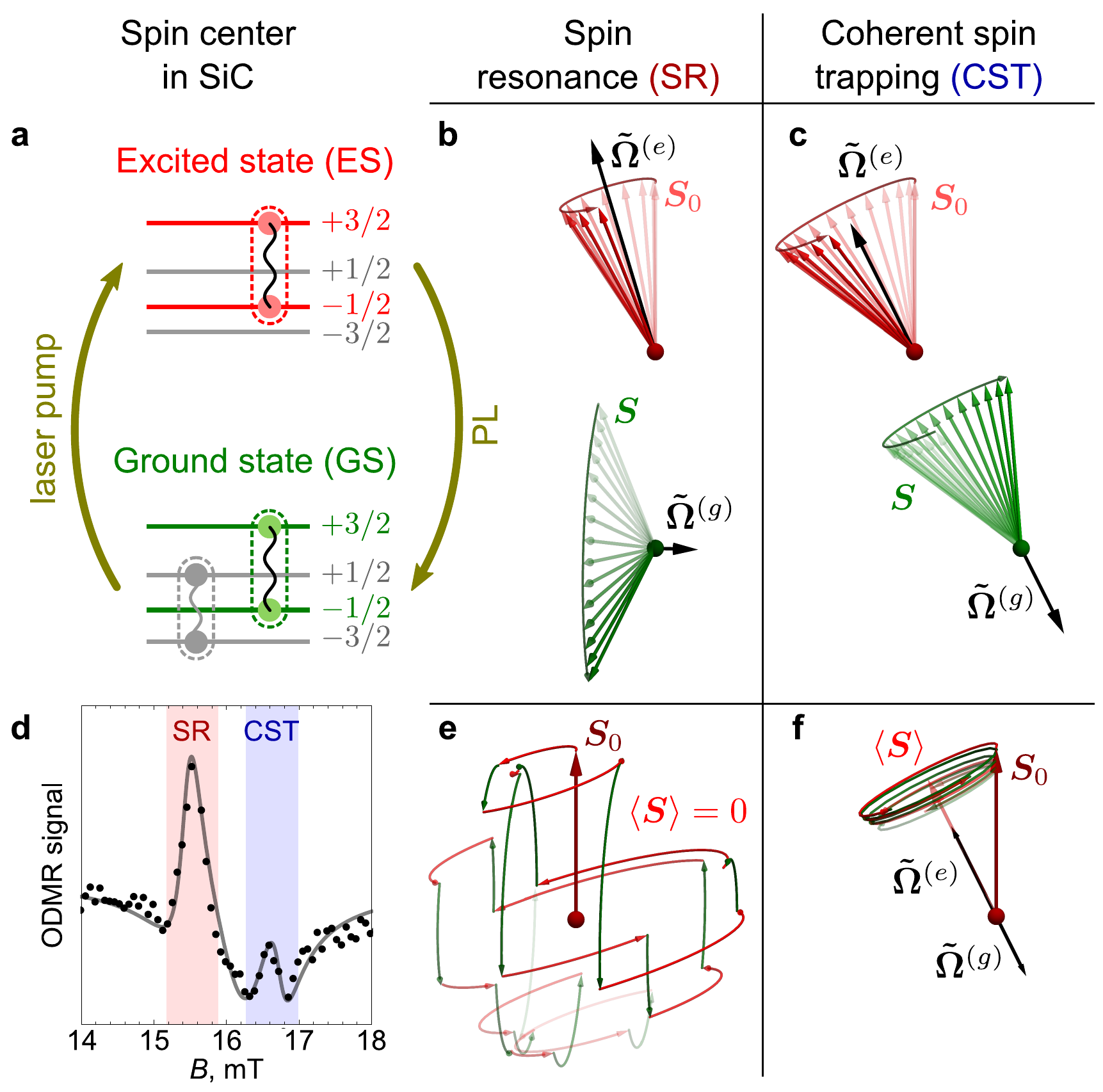}
\caption{\textbf{Comparison of conventional spin resonance (SR) and coherent spin trapping (CST).} \textbf{a,} Electronic level structure of a high-spin ($S= 3/2$) center under optical excitation by a laser pump and decay with photoluminescence (PL). The undulating black curves represent spin transitions driven by an oscillating driving field. \textbf{b} and \textbf{c,} Spin polarization dynamics corresponding to a pair of spin sublevels in the GS and ES under conventional spin resonance (SR) and coherent spin trapping (CST), respectively. \textbf{d,} Example of an optically detected SAR spectrum revealing a pronounced peak due to conventional SR and a quenched resonance with asymmetric Fano-like shape due to CST. Symbols are experimental data and the solid line is a fit as explained in the text. \textbf{e,} In case of SR, the  precession frequencies in GS and ES are not aligned and the spin is rapidly dephased. \textbf{f,} Under CST, the precession frequencies are aligned and the spin projection along this direction is conserved. }
\label{fig1}
\end{figure}

To describe the CST, we consider the electron spin resonance under a static magnetic field ${\bm B} \parallel \hat{\bm y}$ induced by an oscillating field along $\hat{\bm x}$ of frequency $\omega$ and simultaneously driving transitions between two sublevels in the GS and ES of a spin centre (Fig.~\ref{fig1}a). The spin dynamics can be conveniently analysed in the rotating frame approximation, where the reference frame rotates with the driving frequency. In general, the spin precesses around an effective magnetic field $\tilde{\bm \Omega}^{(g)}$ in the GS, and a different effective field $\tilde{\bm \Omega}^{(e)}$ during the time it populates the ES. Both fields consist of a longitudinal {(i.e.,  $\parallel \hat{\bm y}$)} and a transverse {(i.e.,  $\parallel \hat{\bm x}$)} component
\begin{equation}\label{eq:DrivField}
\tilde{\bm \Omega}^{(g,e)}=(  \Delta E^{(g,e)}/\hbar -\omega)\hat{\bm y}+\Omega_R^{(g,e)}\hat{\bm x},
\end{equation}
determined by $\omega$ and the amplitude $\Omega_R^{(g,e)}$  of the driving field as well as by the spin splitting $\Delta E^{(g,e)}$ of the corresponding states.

Under a continuous optical excitation, the centre switches randomly between the GS and ES (Fig.~\ref{fig1}a). This results in a random change between $\tilde{\bm \Omega}^{(g)}$ and $\tilde{\bm \Omega}^{(e)}$, and, therefore, of the spin precession frequency and direction (Fig.~\ref{fig1}b). Here, we assume that the spin is preserved during the optical transitions. The random changes in precession direction lead to a fast spin dephasing (Fig.~\ref{fig1}e), which is analogous to the Dyakonov--Perel spin relaxation mechanism\cite{Dyakonov71}.  The efficient randomization of the spin orientation under such excitation and driving conditions yields a strong spin resonance, as illustrated by the transition masked as SR in Fig.~\ref{fig1}d.  

A different situation emerges if one tunes the frequency of the driving field in such a way that the directions of $\tilde{\bm \Omega}^{(g)}$ and $\tilde{\bm \Omega}^{(e)}$ coincide (Fig.~\ref{fig1}c) -- the CST condition. Then, the spin projection no longer depends on whether the spin is in the GS or ES and, thus, remains conserved for times limited only by intrinsic spin relaxation processes (Fig.~\ref{fig1}f) \cite{Shumilin2015,Shumilin2016,Kubo54}. Such a coherent trapping of the spin polarization leads to a suppression of the spin resonance signal, as illustrated by the transition masked as CST in Fig.~\ref{fig1}d. 

Following our theoretical consideration (Supplementary Information), CST occurs for the driving frequency 
\begin{equation}\label{eq:CST}
\omega_\text{CST} =  \frac{\Delta E^{(g)}\Omega^{(e)}_{R} -\Delta E^{(e)}\Omega^{(g)}_{R}}{\hbar(\Omega^{(e)}_{R} - \Omega^{(g)}_{R})}.
\end{equation}
This condition can always be achieved provided that the driving field amplitudes in the GS and ES are distinct, that is $\Omega_R^{(g)} \neq \Omega_R^{(e)}$. When the spin system is driven by a single radio-frequency (RF) magnetic field $b_\mathrm{rf}$, this requirement can normally not be fulfilled because $\Omega_R^{(g,e)}=\gamma^{(g,e)}b_\mathrm{rf}$ and the gyromagnetic ratios $\gamma^{(g,e)}$ have typically the same value for the GS and ES. In contrast, an oscillating strain field induces spin transitions in the GS and ES with different driving amplitudes due to the different spin-strain interaction constants \cite{Udvarhelyi:2020fq, Anisimov:2016er, Breev:2021dh}, so that CST can be realized experimentally. 
  
\begin{figure*}
\includegraphics[width=0.75\linewidth]{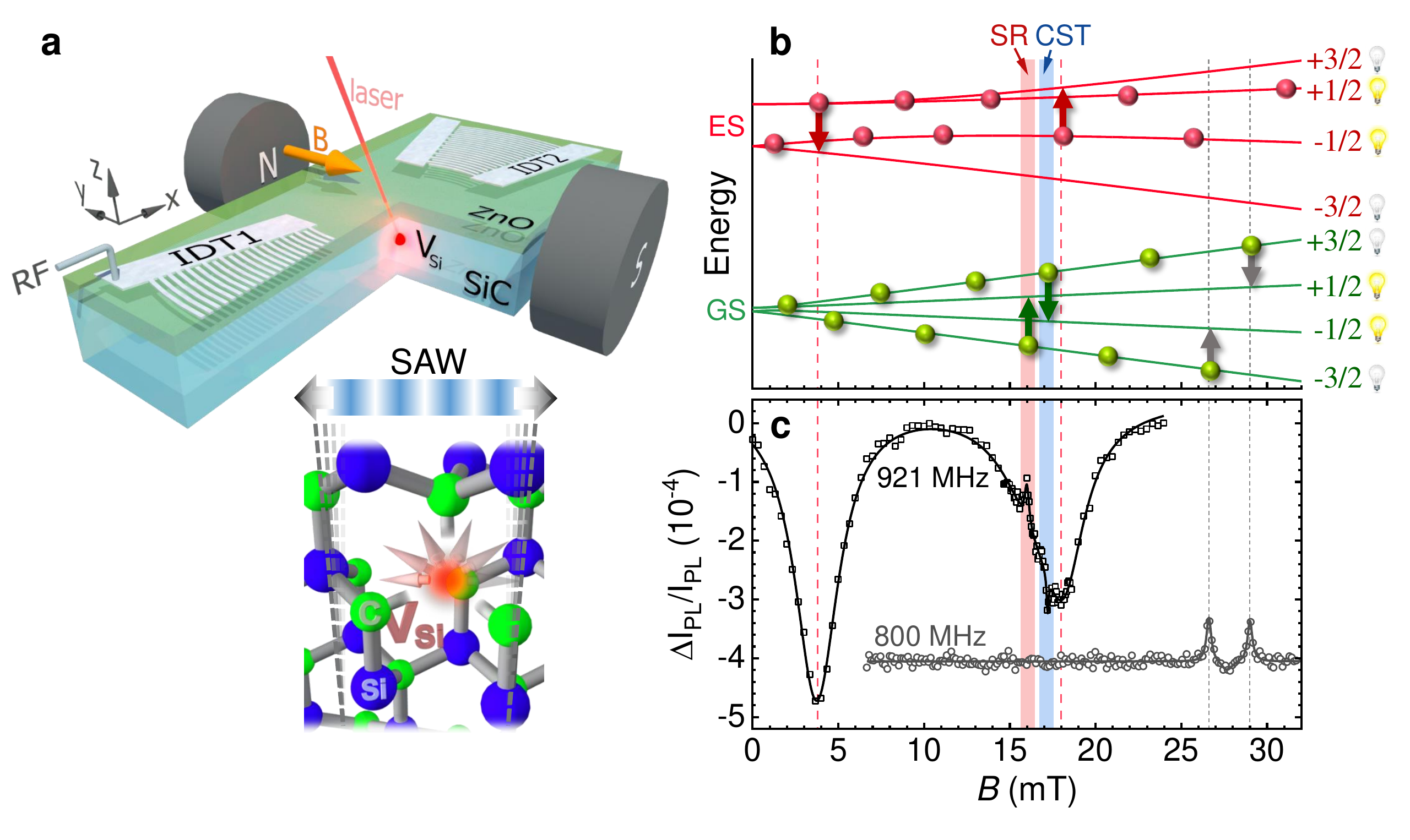}
\caption{\textbf{Ground-state and excited-state spin-acoustic resonances.}  \textbf{a,}  Acoustic resonator consisting of two focusing interdigital transducers (IDTs) exciting a standing SAW oscillating perpendicularly to an in-plane magnetic field $B$. The $\VSi$ centres (inset) created below the 4H-SiC surface are optically pumped by a laser, the excited PL intensity is linked to the occupation of the $\VSi$ spin sublevels. \textbf{b,} Evolution of the spin sublevels in the GS and ES with external magnetic field at 125~K. The green and red vertical arrows indicate the SARs with $\Delta m =\pm 2$ at the SAW frequency of 921~MHz, the grey arrows denote the conventional (i.e., electromagnetically induced) spin transitions measured at 800~MHz. \textbf{c,}  Black squares show a SAR spectrum recorded at 125~K as a function of the external magnetic field for $\wSAW/2\pi=921$~MHz. The solid line is a fit as described in the text. Grey circles display an ODMR signal recorded for an RF signal $\omega/2\pi=800$~MHz applied to the IDT. The spectra are vertically shifted for clarity.} 
\label{fig2}
\end{figure*}

To observe the proposed phenomenon, we make use of the optically detected SAR. Like in the well-known optically detected magnetic resonance (ODMR), SARs are detected by recording the changes in photoluminescence (PL) intensity $\IPL$ when the spin resonance condition is met \cite{HernandezMinguez:2020kv}. We apply this approach to silicon vacancies ($\VSi$) in 4H-SiC with spin $S = 3/2$ (Fig.~\ref{fig2}a). In particular, we concentrate on the so-called V2 $\VSi$ centres, which have well-studied ODMR spectra in external magnetic fields \cite{Kraus:2013di}. An ensemble of $\VSi$ centres is created by proton irradiation at a mean depth of about $2.5~\um$ below the surface (Methods) \cite{Kraus:2017cka}. The $\VSi$ spins are driven by the dynamic strain field of a SAW resonator fabricated on top of the samples (Methods), as  schematically shown in Fig.~\ref{fig2}a. The in-plane magnetic field $\vBext=(0,B_y,0)$, applied perpendicular to the SAW propagation direction, brings the transition frequencies between the spin sublevels of the $\VSi$ centres into resonance with the SAW frequency $\wSAW / 2 \pi$.  

Figure~\ref{fig2}b displays the magnetic field dependence of the spin sublevels in the GS and ES, calculated from the corresponding effective spin Hamiltonians in the uniaxial approximation
\begin{equation}\label{eq:Hamil}
\mathcal{H}^{(g,e)}=D^{(g,e)}\left(S_z^2-\frac{5}{4}\right)+g\mu_B \mathbf{B}\cdot\mathbf{S}.
\end{equation}
\noindent Here, $\mathbf{S}=(S_x,S_y,S_z)$ is the spin-3/2 operator, $\mu_B$ is the Bohr magneton, $g\approx2$ is the  $g$-factor, and $D^{(g,e)}$ are the zero-field splitting constants stemming from the crystal field. Under zero magnetic field, the spin sublevels are split into two Kramer's doublets with a separation of  $2D^{(g)}=70$\,MHz and $2D^{(e)}=430$\,MHz at room temperature between the states with the spin projections $\mS=\pm1/2$ and $\mS=\pm3/2$ on the hexagonal axis  $z$ (perpendicular to the sample plane) \cite{Simin:2016cp}. For strong fields (i.e., $g\mu_B|\Bext|\gg|D^{(e,g)}|$, the spin projection becomes quantized along the magnetic field direction $y$, and all spin sublevels are split (Fig.~\ref{fig2}b). 

Optical excitation from the GS to ES, followed by spin-dependent recombination via the metastable states (MS), leads to the preferential population of the spin states with $\mS=\pm3/2$ in the GS and $\mS=\pm1/2$ in the ES (circles in Fig.~\ref{fig2}b). The latter depends on the excitation conditions \cite{Simin:2016cp, Hoang:2021dx}, a detailed model is presented in the Supplementary Information. Because of the optical spin pumping cycle, the PL intensity is thus stronger for the $\mS=\pm1/2$ states than for the $\mS=\pm3/2$ states \cite{Soykal:2015uw}. Therefore, an increase (decrease) of the PL intensity is expected when the GS (ES) spin transitions  are driven. 

Figure~\ref{fig2}c shows a SAR spectrum (open squares) at $T=125$~K in the magnetic field range expected for  SAR transitions with $\Dm=\pm2$ \cite{Poshakinskiy:2019bi, HernandezMinguez:2020kv}. Under excitation by a $\wSAW/2\pi=921$~MHz SAW (see Supplemental Information for the characterization of the acoustic resonator), we observe two broad dips at $B = 4 \, \mathrm{mT}$ and $B = 18 \, \mathrm{mT}$. They are associated with the ES spin transitions ($+1/2 \rightarrow -3/2$) and ($-1/2 \rightarrow +3/2$), respectively (Fig.~\ref{fig2}b). In addition, the spectrum shows a narrow peak at $B \approx 16 \, \mathrm{mT}$ and a narrow dip at $B \approx 17 \, \mathrm{mT}$. The narrow peak is attributed to the GS spin transition ($-3/2 \rightarrow +1/2$). The narrow dip, in contrast, arises from the interference between the GS and ES ($+3/2 \leftrightarrow -1/2$) transitions leading to the CST. Taking into account that the area of the broad peaks is $\sim 50$ times larger than the area of the narrow peak and dip, and the fact that the spin stays in the ES during much shorter times than in the GS, we estimate the interaction of the SAW with the ES to be about two orders or magnitude stronger than with the GS.

To verify the acoustic nature of the spin transitions, we carried out conventional ODMR experiments under an RF field ($\omega/ 2 \pi = 800$~MHz) below the resonance frequency of the acoustic resonator and, therefore, in the absence of an acoustic field (open circles in Fig.~\ref{fig2}c). The changes in PL contrast only show the $\Dm=\pm1$ spin resonances corresponding to ODMR purely driven by the RF magnetic field, thus confirming that the peak and dips associated with the $\Dm=\pm2$ spin transitions can only be excited by the dynamic SAW strain field.

\begin{figure}
\includegraphics[width=0.9\linewidth]{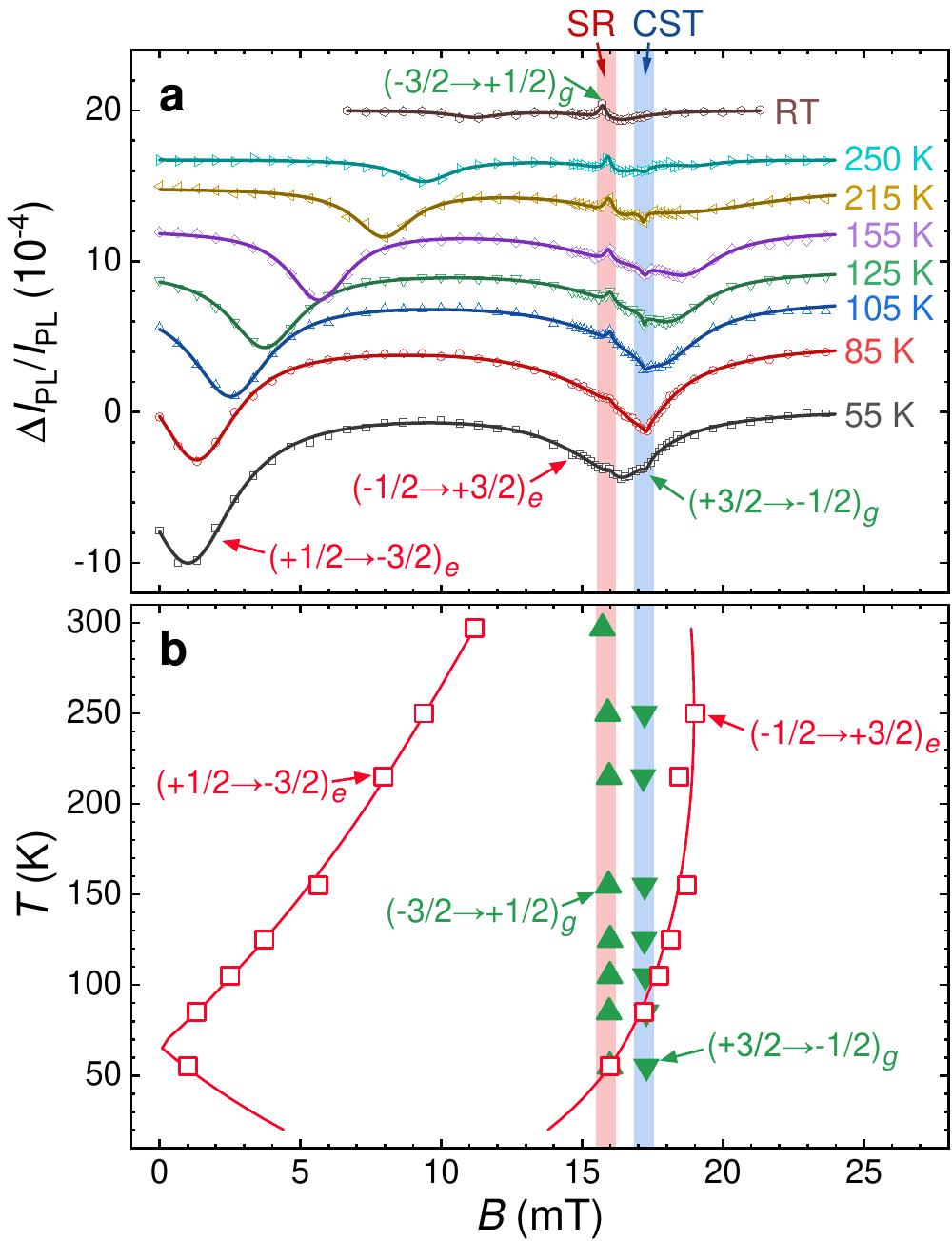}
\caption{\textbf{Temperature dependence of the spin-acoustic resonance.} \textbf{a,}  SAR spectra for different temperatures. The solid curves are fits to Eq.~(\ref{eq:Fano}). The data are vertically shifted for clarity. \textbf{b,} Solid triangles and open squares show the magnetic field positions $B_j$ of the $\Dm=\pm2$ GS and ES spin transitions, respectively, obtained from the fits. Calculations for the ES transitions are represented by the red lines.}
\label{fig3}
\end{figure}

The most remarkable (and unexpected) feature in the SAR spectrum of Fig.~\ref{fig2}c is the negative sign of the GS ($+3/2 \rightarrow -1/2$) spin transition at $B\approx17 \, \mathrm{mT}$ (note that, according to preferential population of the spin states, the sign  of the resonance should be positive, as for the spin transition at $B\approx 16$ mT). Detailed studies presented below reveal that it has even a more complex, Fano-like shape. Such a behaviour is a fingerprint of CST. To further confirm this assumption, we measured SAR spectra at different temperatures (Fig.~\ref{fig3}). While the GS zero-field splitting $2D^{(g)}$ is almost temperature independent, the ES zero-field splitting $2D^{(e)}$ increases at a rate of $2.1$~MHz/K with reducing temperature~\cite{Anisimov:2016er}, thus shifting the spin resonances of the ES towards lower magnetic fields (Fig.~\ref{fig3}a). 

We fit the experimental curves by a sum of Fano-like resonance functions \cite{Limonov2017}
\begin{align}\label{eq:Fano}
\Delta\IPL/\IPL = \sum_j \frac{A_j \delta B_j^2 + Q_j  (B-B_j) \delta B_j}{(B-B_j)^2 + \delta B_j^2} \,, 
\end{align}
where the summation goes over all the GS and ES spin transitions. Here, $B_j$ denote the magnetic field positions of the resonances, $\delta B_j$ are their widths, $A_j$ and $Q_j$ are the amplitudes of the symmetric and anti-symmetric parts of the resonances. The values of $B_j$ are depicted in Fig.~\ref{fig3}b, together with the theoretical temperature dependences for the ES resonances (the red lines in Fig.~\ref{fig3}b) obtained from the spin Hamiltonian in Eq.~(\ref{eq:Hamil}) and the known temperature dependence of $2D^{(e)}$ \cite{Anisimov:2016er}. Importantly, the temperature variation allows direct monitoring of the CST as a function of the detuning between the ES spin transition and its GS counterpart. In parallel, the other ES spin resonance ($+1/2 \rightarrow -3/2$) is strongly detuned from its GS counterpart for all temperatures. The latter behaves as a normal spin resonance and, hence, can be used as a reference. 

\begin{figure}
\includegraphics[width=0.9\linewidth]{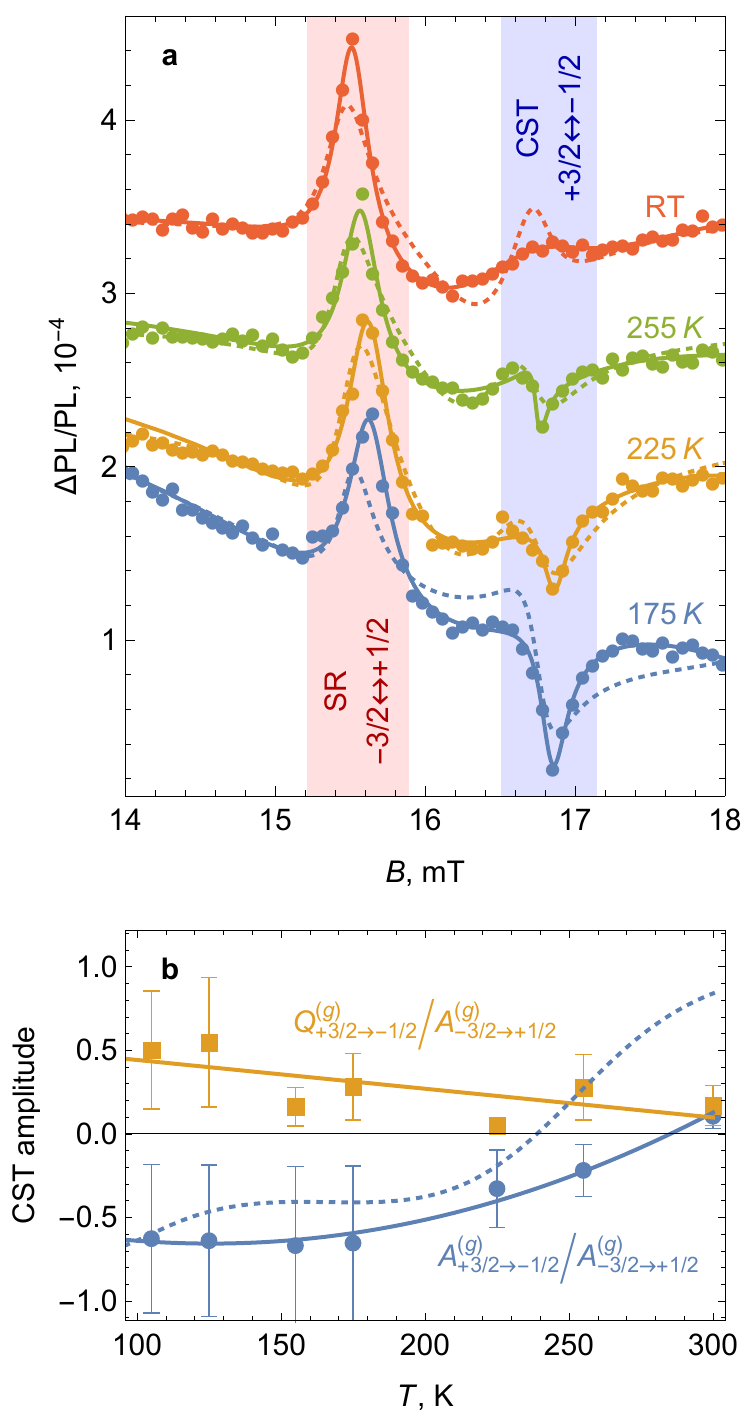}
\caption{\textbf{Coherent spin trapping for different detunings.} \textbf{a,} Detailed measurement of the $\Dm=\pm2$ GS SAR spectra at representative temperatures. The solid lines are fits to Eq.~(\ref{eq:Fano}). The dashed lines are calculations based on our microscopic model. The data are vertically shifted for clarity. 
\textbf{b,} Parameters of the  ($-3/2\rightarrow +1/2$)  resonance as a function of temperature. Amplitudes of the symmetric and anti-symmetric parts of the resonance line are shown. The values are normalized by the amplitude of  the ($+3/2\rightarrow -1/2$) spin resonance. The solid lines are a guide for eye, the dashed line is calculated after the model. 
}
\label{fig4}
\end{figure}

Detailed temperature dependences of the GS $\Dm=\pm2$ spin resonances are presented in Fig.~\ref{fig4}a. At room temperature (RT), both spin resonances appear as positive peaks in the SAR spectrum, but the $(+3/2\rightarrow -1/2)$ resonance at $17$~mT is almost suppressed due to the CST mechanism. As the temperature decreases and the detuning between the ($-1/2\leftrightarrow+3/2$) spin resonances in the GS and ES is reduced, the spin transition at $17$~mT becomes asymmetric and eventually changes its sign. The fits by Eq.~(\ref{eq:Fano}) are in very good agreement with our experimental data (solid lines in Fig.~\ref{fig4}a). We find that $Q_j=0$ for all resonances except for the GS $(+3/2\rightarrow -1/2)$ spin transition, indicating its asymmetric Fano-like shape. We attribute this exceptional feature to the interference with the closely located ES $(-1/2\rightarrow +3/2)$ spin resonance leading to the CST. Figure~\ref{fig4}(b) shows the temperature dependence of the symmetric and anti-symmetric parts of the resonance at $17$~mT,  $A_{+3/2\rightarrow -1/2}^{(g)}$ (blue circles) and $Q_{+3/2\rightarrow -1/2}^{(g)}$ (yellow squares). We normalized them by the amplitude of the other GS resonance $A_{-3/2\rightarrow +1/2}^{(g)}$, which  remains always positive since its counterpart ES $(+1/2\rightarrow -3/2)$  is  far detuned. The negative amplitude of $A_{+3/2\rightarrow -1/2}^{(g)}$ at low temperatures originates from the interference of the ES and GS $(+3/2\leftrightarrow -1/2)$ transitions. As the temperature increases, its value goes to zero at around 280\,K and then becomes slightly positive, reflecting the suppression of the interference due to the increase of the detuning between the resonances and the speedup of the decoherence processes at high temperatures. The anti-symmetric contribution to the resonance is always present, but it becomes comparable to the symmetric one at $T\approx 250$\,K, giving rise to the strongly asymmetric lineshape. 

To model the magnetic field dependencies, we develop a microscopic model, which includes the spin sublevels in the GS and ES, together with an intermediate metastable state \cite{Soykal:2015uw} (Supplementary Information). We take into account spin-conserving optical transitions between the GS and ES, and a spin-dependent transition rate to the metastable state, which enables spin polarization under optical pumping and explains the dependence of the PL intensity on the degree of spin polarization. Our model predicts that far-detuned resonances between the same spin sublevels in the GS and ES should lead to the positive and negative sign of the SAR, respectively (Supplementary Information), in agreement with  the GS $(-3/2\rightarrow +1/2)$ and ES $(+1/2\rightarrow -3/2)$ resonances in the experimental spectra. However, if the detuning becomes comparable to the width of the ES reonance, then the GS resonance takes an asymmetric lineshape and eventually flips its sign. The dashed lines in Fig.~\ref{fig4}a,b display the calculated SAR spectra and the resonance amplitudes following this model. They show a good quantitative agreement with our experimental data and reveal all essential features of the CST mechanism. Another prediction of the model is a drastic decrease of the GS resonance width in the CST conditions due to the increased spin lifetime. However, in our experiment the change of the homogeneous linewidth is masked by a stronger inhomogeneous broadening, originating from fluctuations of the zero-field splittings and nuclear fields. In case of smaller inhomogeneous broadening, sharper lines with more pronounced asymmetry are expected (Supplementary Information). 

In conclusion, we have demonstrated that SAWs can efficiently control transitions in atomic-scale spin centres both in the GS and ES. Their simultaneous driving with the same SAW field leads to the coherent trapping of the spin polarization along a well-defined direction. It manifests itself as a suppression of the spin relaxation compared to a canonical spin resonance under the same conditions. We have developed a microscopic model for the acoustically induced CST, which shows a good quantitative agreement with the experimental data. Acoustic excitation opens the way for the coherent manipulation of GS and ES spins.
The use of a  single SAW resonator establishes a single spin conservation axis and, in addition, limits the detuning range and, therefore, the temperature window over which CST can be observed. These limitations can be overcome by multi-color (i.e., multi-frequency) SAW resonators orthogonally oriented for the excitation of coherent GS and ES transitions along arbitrary directions. In combination with the double RF-optical resonance \cite{Riedel:2012jq} and fast reconfigurable quantum emitters \cite{Lukin:2020jb}, our approach can be extended for the control of individual spin qubits with coherent acoustic and optical fields. The acoustic driving approach is thus promising for the implementation of quantum transducers \cite{Schuetz:2015dx}, mechanical cooling \cite{MacQuarrie:2017hl}, coherent sensing \cite{Arcizet:2011cg, Barfuss:2015hv, Kolkowitz:2012iw, Poshakinskiy:2019bi} as well as photon \cite{Machielse:2019bt} or phonon \cite{Lemonde:2018jx} networks; all these represent milestones towards on-chip quantum information processing on different material platforms.

\section*{Acknowledgements}

The authors would like to thank S.~Meister and S.~Rauwerdink for technical support in the fabrication of the acoustic resonators, S.~A.~Tarasenko, M.~M.~Glazov, and M.~Helm for discussions and critical questions, and J. L\"ahnemann for a critical reading of the manuscript. A.V.P. acknowledges the support from the Russian Science Foundation (project 19-72-00080), Russian President Grant No. MK-4191.2021.1.2, and the Foundation ``BASIS''. G.V.A. acknowledges the support from  the German Research Foundation (DFG) under Grant  AS 310/9-1. Support from the Ion Beam Center (IBC) at Helmholtz-Zentrum Dresden-Rossendorf (HZDR) is gratefully acknowledged for the proton irradiation.

\section*{Author contributions}
AHM supervised the fabrication of the acoustic resonators and performed the optically detected spin-acoustic resonance experiments. AVP developed the theoretical model and analysed the experimental data. MH calculated irradiation parameters and characterized the $\VSi$ centers. AHM, AVP, PVS and GVA discussed the results and contributed to the production of the manuscript.

\section*{Methods}

\subsection*{Sample fabrication}
The $\VSi$ centers were created in a $10\times10$~mm$^2$ semi-insulating 4H-SiC substrate by the irradiation with protons with an energy of 375~keV and a fluence of $10^{15}$~cm$^{-2}$. After irradiation, the SiC substrate was coated with a 35-nm-thick SiO$_2$ layer followed by a 700-nm-thick ZnO piezoelectric film using radio-frequency (RF) magnetron sputtering. Acoustic resonators defined by a pair of focusing interdigital transducers (IDTs) were then patterned on the surface of the ZnO film by electron beam lithography and metal evaporation. Each IDT consists of 80 aluminium finger pairs for the excitation/detection of SAWs with a wavelength $\lSAW=6~\um$. An additional acoustic Bragg reflector consisting of 40 finger pairs was placed on the IDT's back side. The finger curvature and separation between the opposite IDTs ($\approx 120~\um$) are designed to focus the SAW beam at the center of the resonator.  

\subsection*{Measurements}
The optically detected spin-acoustic resonance experiments were performed in a confocal micro-photoluminescence ($\mu$-PL) setup. The sample was placed in a cold-finger cryostat equipped with a window for optical access and coaxial connections for the application of the RF signals to the IDTs. The spin transitions in the $\VSi$ centres were tuned to the frequency of the SAW by an in-plane magnetic field applied using an electromagnet. The $\VSi$ centres were optically excited by a 780~nm laser beam focused onto a spot size of $10~\mu$m by a $20\times$ objective with numerical aperture NA=0.4. The $\VSi$ PL band centred around 917~nm was collected by the same objective, spectrally separated from the reflected laser beam using an 805~nm dichroic mirror and an 850~nm long-pass filter, and detected by a silicon photodiode. The output signal of the photodiode was locked-in to the frequency of the amplitude modulation of the RF signal applied to the IDT.


\end{document}


\title{Acoustically induced coherent spin trapping}
\author{A.~Hern\'{a}ndez-M\'{i}nguez}
\email{alberto.h.minguez@pdi-berlin.de}
\affiliation{Paul-Drude-Institut f\"{u}r Festk\"{o}rperelektronik, Leibniz-Institut im Forschungsverbund Berlin e.V., Hausvogteiplatz 5-7, 10117 Berlin, Germany}
\author{A.~V.~Poshakinskiy}
\affiliation{Ioffe Physical-Technical Institute, Russian Academy of Sciences, 194021 St.~Petersburg, Russia}
\author{M.~Hollenbach}
\affiliation{Helmholtz-Zentrum Dresden-Rossendorf, Institute of Ion Beam Physics and Materials Research, Bautzner Landstrasse 400, 01328 Dresden, Germany}
\affiliation{Technische Universit\"{a}t Dresden, 01062 Dresden, Germany}
\author{P.~V.~Santos}
\affiliation{Paul-Drude-Institut f\"{u}r Festk\"{o}rperelektronik, Leibniz-Institut im Forschungsverbund Berlin e.V., Hausvogteiplatz 5-7, 10117 Berlin, Germany}
\author{G.~V.~Astakhov}
\affiliation{Helmholtz-Zentrum Dresden-Rossendorf, Institute of Ion Beam Physics and Materials Research, Bautzner Landstrasse 400, 01328 Dresden, Germany}


\begin{abstract}
This Supplemental Material contains information about:
\begin{itemize}
\item Characterization of the acoustic resonator
\item Theory of coherent spin trapping
\item Optically detected spin trapping
\item Fit of experimental ODMR spectra
\end{itemize}
\end{abstract}

\maketitle

\section{Characterization of the acoustic resonator}\label{sec_IDTs}

\begin{figure}[h]
\includegraphics[width=0.8\linewidth]{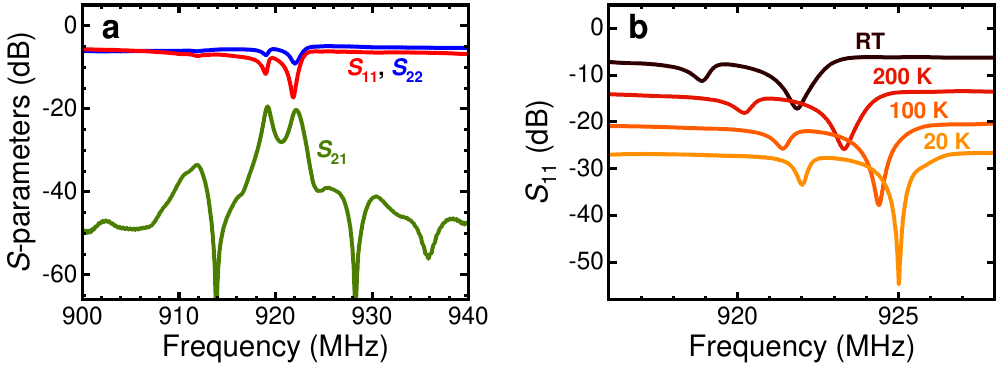}
\caption{\textbf{a,} Room-temperature power scattering parameters of the acoustic resonator as a function of the applied RF frequency. $\sleft$ and $\sright$ are the power reflection coefficients of IDT1 and IDT2, respectively, while $\st$ is the power transmission coefficient. The $\st$ spectrum is time-gated to remove the electromagnetic cross-talk. \textbf{b,} Temperature dependence of the $\sleft$ coefficient. The curves are vertically shifted for clarity.}
\label{fig_IDTs}
\end{figure}

Figure~\ref{fig_IDTs}a displays the RF power scattering ($S$) parameters of IDT1 and IDT2 in the acoustic resonator, measured at room temperature with a vector network analyser. The reflection coefficients ($\sleft$, $\sright$) show a series of dips within the resonance band of the IDTs, accompanied by a series of peaks in the transmission coefficient ($\st$). Such dips/peaks correspond to the excitation of the Rayleigh SAW modes in the resonator. Figure~\ref{fig_IDTs}b displays the temperature dependence of the $\sleft$ reflection coefficient. The SAW resonances move towards higher frequencies as the temperature decreases. This blue shift was taken into account in the measurements as a function of temperature in the manuscript.

\section{Theory of coherent spin trapping}

To introduce the concept of coherent spin trapping (CST), we first consider a simplified model that is limited to one ground state (GS) and one excited state (ES). The full model used to describe the experimental optically detected magnetic resonance (ODMR) spectra features additionally an intermediate metastable state (MS) that is necessary to explain the optical pump and read-out of the spin polarization, see Sec.~\ref{sec:MS}.
 We limit the consideration to just two spin sublevels in the GS and two spin sublevels in the ES. Then, the state of the system is described by the spin density matrices $2\times 2$ of the GS and ES, 
\begin{align}
\rho_{g} &= \frac12 N_g + \bm S_g \cdot \bm \sigma, \\
\rho_{e} &= \frac12 N_e + \bm S_e \cdot \bm \sigma ,
\end{align}
where $N_g$ and $N_e$ are the occupancies of the GS and ES, while $\bm S_g$ and $\bm S_e$ are the pseudospins of the GS and ES. 
Optical transitions between GS and ES are supposed to be incoherent and described by the rate equation. Within the GS and the ES, the evolution of the spin is assumed to be coherent; when the system goes from the GS to the ES and vice versa, spin coherence is assumed to be preserved. The dynamical equations read:
\begin{align}
\frac{d\rho_g}{dt} = \frac{\rmi}{\hbar} [\rho_g, H_g] + \Gamma \rho_e - P \rho_g, \\ 
\frac{d\rho_e}{dt} = \frac{\rmi}{\hbar} [\rho_e, H_e] - \Gamma \rho_e + P \rho_g,
\end{align}
where
\begin{align}
H_{g,e} = (\hbar/2) \bm\Omega^{(g,e)}(t) \cdot \bm\sigma
\end{align}
are the Hamiltonians in the GS and ES, $P$ is the optical pumping rate, and $\Gamma$ is the relaxation rate of the ES. 
The spin and population dynamics are described by the decoupled equations:
\begin{align}
&\frac{dN_g}{dt} =  \Gamma N_e - P N_g, \\ 
&\frac{dN_e}{dt} = - \Gamma N_e + P N_g,
\end{align}
and
\begin{align}
\frac{d\bm S_g}{dt} &= \bm\Omega^{(g)}(t) \times \bm S_g+ \Gamma  \bm S_e - P  \bm S_g -\gamma  \bm S_g + \Sigma \hat{\bm z},
\label{eq:dSg}\\
\frac{d\bm S_e}{dt} &=\bm\Omega^{(e)}(t) \times \bm S_e- \Gamma  \bm S_e + P \bm S_g -\gamma  \bm S_e,\label{eq:dSe}
\end{align}
where we introduced the spin relaxation rate $\gamma$ and the phenomenological spin pumping rate $\Sigma$ (the origin of the spin pumping is the relaxation via a MS, as described in Sec.~\ref{sec:MS}). 

To calculate the spin resonance spectra, we take
\begin{align}
\bm\Omega^{(g,e)}(t) = \Omega^{(g,e)}_{z} \hat{\bm z} + \Omega_{\rm rf}^{(g,e)}(t),
\end{align} where
\begin{align}
\Omega^{(g,e)}_{z} = \Delta E^{(g,e)}/\hbar
\end{align}
are the Zeeman splittings of the considered spin state pairs in the GS and ES, and
\begin{align}
\Omega_{\rm rf}^{(g,e)}(t) = \text{Re\,}\Omega_R^{(g,e)} (\hat{\bm x} + \rmi \hat{\bm y}) \e^{-\rmi\omega t} 
\end{align}
is the driving field. For simplicity, we assume that the latter  is circularly polarized, and switch to the reference frame rotating with its frequency $\omega$.  There, the spin dynamics is governed by the static effective field
\begin{align}\label{eq:Om}
\tilde{\bm\Omega}^{(g,e)} = (\Omega^{(g,e)}_{z}-\omega) \hat{\bm z} + \Omega_R^{(g,e)}\hat{\bm x}.
\end{align}
The normalized spin-resonance (SR) signal is found as
\begin{align}
\text{R} =1-  \gamma ( S^{(0)}_{g,z} + S^{(0)}_{e,z})/\Sigma,
\end{align}
where $S^{(0)}_{g,e}$ are the steady-state solutions of Eqs.~\eqref{eq:dSg}-\eqref{eq:dSe} with $\bm\Omega^{(g,e)}(t)$ replaced by $\tilde{\bm\Omega}^{(g,e)}$, Eq.~\eqref{eq:Om}. 

First, we consider the limit $|\Omega^{(e)}_{z} - \Omega^{(g)}_{z}| \to \infty$, that is, when the spin resonances corresponding to the GS and ES are well separated spectrally. The SR signal in the vicinity of each resonance is then given by
\begin{align}
\text{R}(\omega)_g = \frac{\frac{P \Gamma}{\Gamma+P}\, \frac{\Omega_{R,g}^2}{\gamma}}{ (\Omega_{z}^g-\omega)^2 + P^2 + \frac{P \Gamma}{\Gamma+P}\, \frac{\Omega_{R,g}^2}{\gamma}} \,,\\
\text{R}(\omega)_e = \frac{\frac{P \Gamma}{\Gamma+P}\, \frac{\Omega_{R,g}^2}{\gamma}}{ (\Omega_{z}^e-\omega)^2 + \Gamma^2 +\frac{P \Gamma}{\Gamma+P}\, \frac{\Omega_{R,g}^2}{\gamma}} \,,
\end{align}
where we have assumed $\Gamma \sim P \gg \gamma$. When the resonances corresponding to the GS and ES overlap, then the SR signal reads
\begin{align}
\text{R} = \frac{\Gamma P}{\gamma(\Gamma + P)} \,\, \frac{[(\Omega^{(g)}_{z}-\omega) \Omega_R^{(e)} - (\Omega^{(e)}_{z}-\omega) \Omega_R^{(g)}]^2}{[\Gamma(\Omega^{(g)}_{z}-\omega) + P (\Omega^{(e)}_{z}-\omega) ]^2 + (\Omega^{(g)}_{z}-\omega)^2 (\Omega^{(e)}_{z}-\omega)^2+ [(\Omega^{(g)}_{z}-\omega) \Omega_R^{(e)} - (\Omega^{(e)}_{z}-\omega) \Omega_R^{(g)}]^2}.
\end{align}
Note that the SR signal vanishes if
\begin{align}
\frac{\Omega^{(g)}_{z}-\omega}{\Omega^{(e)}_{z}-\omega} = \frac{\Omega_R^{(g)}}{\Omega_R^{(e)}},
\end{align}
i.e., when the total precession frequencies in the GS and ES are collinear. This is realized at 
\begin{align}
\omega_\text{CST} = \frac{\Omega^{(g)}_{z}\Omega^{(e)}_{R} - \Omega^{(e)}_{z}\Omega^{(g)}_{R}}{\Omega^{(e)}_{R} - \Omega^{(g)}_{R}} .
\end{align}
Under this condition, the spin precesses around the same direction in the GS and ES. As a consequence, the spin projection along this direction does not dephase under the random switching between the GS and ES, but conserves its coherence on the long time scale $1/\gamma$. This situation is what we define as CST. Away from the CST condition, the precession frequency and direction change continuously as the centre switches randomly between the GS and ES. This causes the spin to dephase with the rate $\sim [(\Omega^{(e)}_{z} - \Omega^{(g)}_{z})^2+(\Omega^{(e)}_{R} - \Omega^{(g)}_{R})^2]/(\Gamma+P)$, \'a la Dyakonov--Perel spin relaxation mechanism. 

The angle between the trapped spin component and $z$ axis is given by
\begin{align}
\theta_\text{CST} = \frac{\Omega^{(e)}_{R} - \Omega^{(g)}_{R}}{\Omega^{(e)}_{z} - \Omega^{(g)}_{z}} \,.
\end{align}
The SR signal at CST is given by
\begin{align}
\text{R}(\omega_\text{CST}) = 1- \cos^2 \theta_\text{CST} = \frac{(\Omega^{(e)}_{R} - \Omega^{(g)}_{R})^2}{(\Omega^{(e)}_{z} - \Omega^{(g)}_{z})^2+(\Omega^{(e)}_{R} - \Omega^{(g)}_{R})^2} \,. 
\end{align}
At small Rabi frequencies, this value is small, so CST is revealed as a sharp spectral dip.


\section{Optically detected spin trapping}\label{sec:MS}

\begin{figure}[b]
\includegraphics[width=0.4\textwidth]{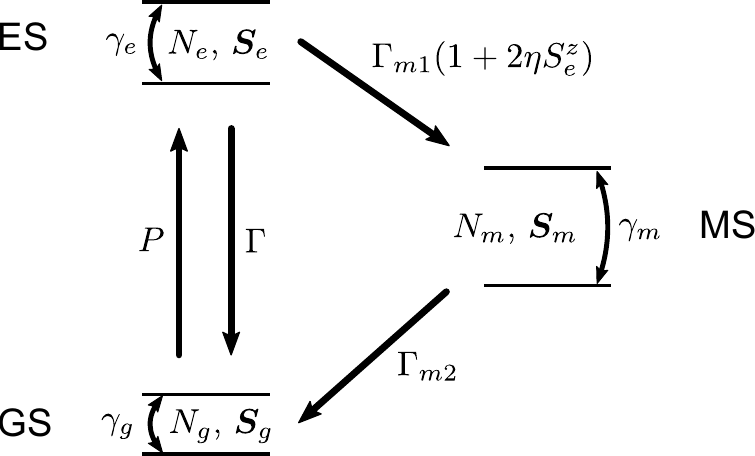}
\caption{The three-level model of the spin centre. }
\label{fig:SM}
\end{figure}

To describe the manifestation of the CST in the spectra of  the ODMR, we consider a model that accounts for the GS, ES and intermediate MS, see Fig.~\ref{fig:SM}. Only two spin sublevels of each level are considered (pseudospin 1/2). To explain optical orientation (appearance of polarization under non-resonant optical pumping) and read-out (variation of the photoluminescence (PL) intensity depending of the spin polarization), the ES to MS relaxation is assumed to be spin dependent (parameter $\eta$). The model is described by the equation set for the spin-density matrices $\rho_g$, $\rho_e$, and $\rho_m$ for GS, ES, and MS: 
\begin{align}
&\frac{d\rho_e}{dt} = \frac{\rmi}{\hbar} [\rho_e, H_e] - \Gamma \rho_e -\Gamma_{m1} \{(1+\eta \sigma_z),  \rho_e\}+ P \rho_g \,,\\
&\frac{d\rho_g}{dt} = \frac{\rmi}{\hbar} [\rho_g, H_g] + \Gamma \rho_e - P \rho_g + \Gamma_{m2} \rho_m \,,\\ 
&\frac{d\rho_m}{dt} = \frac{\rmi}{\hbar} [\rho_m, H_m] + \Gamma_{m1} \{(1+\eta \sigma_z),  \rho_e\} - \Gamma_{m2} \rho_m \,.
\end{align}
This yields the coupled equations for the populations
\begin{align}
&\frac{dN_e}{dt} =  - \Gamma N_e -\Gamma_{m1} (N_e+2\eta S_{e}^{z})+ P N_g \,,\\
&\frac{dN_g}{dt} = \Gamma N_e - P N_g + \Gamma_{m2} N_m \,,\\ 
&\frac{dN_m}{dt} =  \Gamma_{m1} (N_e+2\eta S_{e}^{z}) - \Gamma_{m2} N_m \,,
\end{align}
and spins
\begin{align}
&\frac{d\bm S_e}{dt} = \bm\Omega_e \times \bm S_e -\gamma_e \bm S_e - \Gamma \bm S_e -\Gamma_{m1} (\bm S_e+\tfrac12 \eta N_{e}\bm e^{z})+ P \bm S_g \,,\\
&\frac{d\bm S_g}{dt} = \bm\Omega_g \times \bm S_g  -\gamma_g \bm S_g +\Gamma \bm S_e - P \bm S_g + \Gamma_{m2} \bm S_m \,,\\ 
&\frac{d\bm S_m}{dt} =  \bm\Omega_m \times \bm S_m -\gamma_m \bm S_m + \Gamma_{m1} (\bm S_e+\tfrac12 \eta N_{e}\bm e^{z}) - \Gamma_{m2} \bm S_m \,.
\end{align}
The PL intensity is given by
\begin{align}
I_\text{PL} = \Gamma N_e\,.
\end{align}
We assume $\eta \ll 1$ and evaluate PL up to $\eta^2$. Neglecting $\eta$, the steady-state ES population reads
\begin{align}
N_e^{(0)} = \frac{\Gamma_{m2} P}{\Gamma_{m1} (\Gamma_{m2}+P)+\Gamma_{m2} (\Gamma +P)} \,.
\end{align}
Spin-dependent relaxation from ES to MS leads to the generation of the spin in the ES and MS which is described by
$\bm G_e = -\tfrac12 \eta \Gamma_{m1} N_e^{(0)} \hat{\bm z} $, $\bm G_g = 0$,  $\bm G_m = \tfrac12 \eta \Gamma_{m1} N_e^{(0)} \hat{\bm z} $,
and the following steady-state spin polarizations (up to terms linear in $\eta$):
\begin{align}
\left(\begin{array}{c} S_e^z \\ S_g^z \\ S_m^z \end{array}\right) &= \left(\begin{array}{c}  -(\gamma_g \Gamma_{m2} +\gamma_m P + \gamma_g \gamma_m) \\ \gamma_e \Gamma_{m2} -\gamma_m \Gamma \\ \gamma_g \Gamma +\gamma_e P + \gamma_g \gamma_e\end{array}\right) \nonumber\\
&\times\frac{\tfrac12\eta N_e^{(0)} \Gamma_{m1} }{\gamma_g \gamma_m (\Gamma +\gamma_e+\Gamma_{m1})+\gamma_g \Gamma_{m2} (\Gamma +\gamma_e+\Gamma_{m1})+\gamma_m P (\gamma_e+\Gamma_{m1})+\gamma_e \Gamma_{m2} P} \,.
\end{align}

Note that the GS and ES have opposite spin orientation if
\begin{align}\label{eq:cond}
\frac{\Gamma_{m2}}{\gamma_m} > \frac{\Gamma}{\gamma_e}\,.
\end{align}
This condition means that the spin coming to the GS from the MS ($\propto \Gamma_{m2}/\gamma_m$) dominates over the spin coming to the GS from the ES ($\propto \Gamma/\gamma_e$). 
If condition \eqref{eq:cond} is fulfilled, the ODMR signal is negative for the ES and positive for the GS (as observed in the experiment). If the condition \eqref{eq:cond} is not fulfilled, then the ODMR signal is negative for both ES and GS. 
For simplicity, we assume in the following that $\gamma_m = 0$, thus fulfilling the condition \eqref{eq:cond}. We also assume that the MS is far off-resonance, $\Omega_m^z \to \infty$, which corresponds to $S_m^x,S_m^y = 0$. 
%
%
Then, up to terms quadratic in $\Omega_R^{(g)}$ and $\Omega_R^{(e)}$, the ODMR signal reads
\begin{align}\label{eq:dPL}
&\Delta\text{PL}/\text{PL} \propto \\\nonumber
&\frac{\eta^2}{{(\gamma_g (\Gamma +\gamma_e)+\gamma_e P)^2 \left(\Delta\omega_e^2 \left(\Delta\omega_g^2+(\gamma_g+W_g+P)^2\right)+2 \Delta\omega_e \Delta\omega_g \Gamma  P+\Delta\omega_g^2 (\Gamma +\gamma_e)^2+((\Gamma +\gamma_e) (\gamma_g+W_g)+\gamma_e P)^2\right)}}\\\nonumber
&\times \Big[  \gamma_e P \Omega_R^{(g)2} \left(\Delta\omega_e^2 (\gamma_g+W_g+P)+(\Gamma +\gamma_e) ((\Gamma +\gamma_e) (\gamma_g+W_g)+\gamma_e P)\right)
\\\nonumber
&+P \Omega_R^{(g)}\Omega_R^{(e)} (\gamma_e (\gamma_g+P)-\Gamma  \gamma_g) (-\Delta\omega_e \Delta\omega_g+(\Gamma +\gamma_e) (\gamma_g+W_g)+\gamma_e P)
\\\nonumber
&-\gamma_g \Omega_R^{(e)2} (\gamma_g+P) \left(\Delta\omega_g^2 (\Gamma +\gamma_e)+(\gamma_g+W_g+P) ((\Gamma +\gamma_e) (\gamma_g+W_g)+\gamma_e P)\right)\Big] \,,
\end{align}
where $\Delta\omega_{g,e} = \Omega^{(g,e)}_z-\omega$ and we supposed $\Gamma_{m1},\Gamma_{m2} \to 0$ for simplicity. We also introduced the inhomogeneous GS resonance broadening $W_g$.

Note that the width of the GS resonance in the ODMR spectrum reads
\begin{align}
\Gamma_g = \gamma_g  +P\left(1-\frac{\Gamma(\Gamma+\Gamma_{m1}+\gamma_e)}{(\Omega^{(e)}_z-\Omega^{(g)}_z)^2+(\Gamma+\Gamma_{m1}+\gamma_e)^2}\right) +W_g \,.
\end{align} 
When the GS and ES resonances are well separated, $|\Omega^{(e)}_z-\Omega^{(g)}_z| \to \infty$, the width of the GS resonance is
\begin{align}
\Gamma_g^{(0)} = \gamma_g  +P +W_g \,.
\end{align} 
As the two resonances come close to each other, the width decreases due to the CST and, for $\Omega^{(e)}_z=\Omega^{(g)}_z$, it reduces to
\begin{align}
\Gamma_g^{\text{(CST)}} =  \gamma_g  +\frac{\Gamma_{m1}+\gamma_e}{\Gamma+\Gamma_{m1}+\gamma_e}\,P +W_g\,,
\end{align} 
i.e., the term corresponding to the pump-induced widening $P$ is suppressed.

\section{Fit of experimental ODMR spectra}

\begin{figure}[t]
\includegraphics[width=0.5\textwidth]{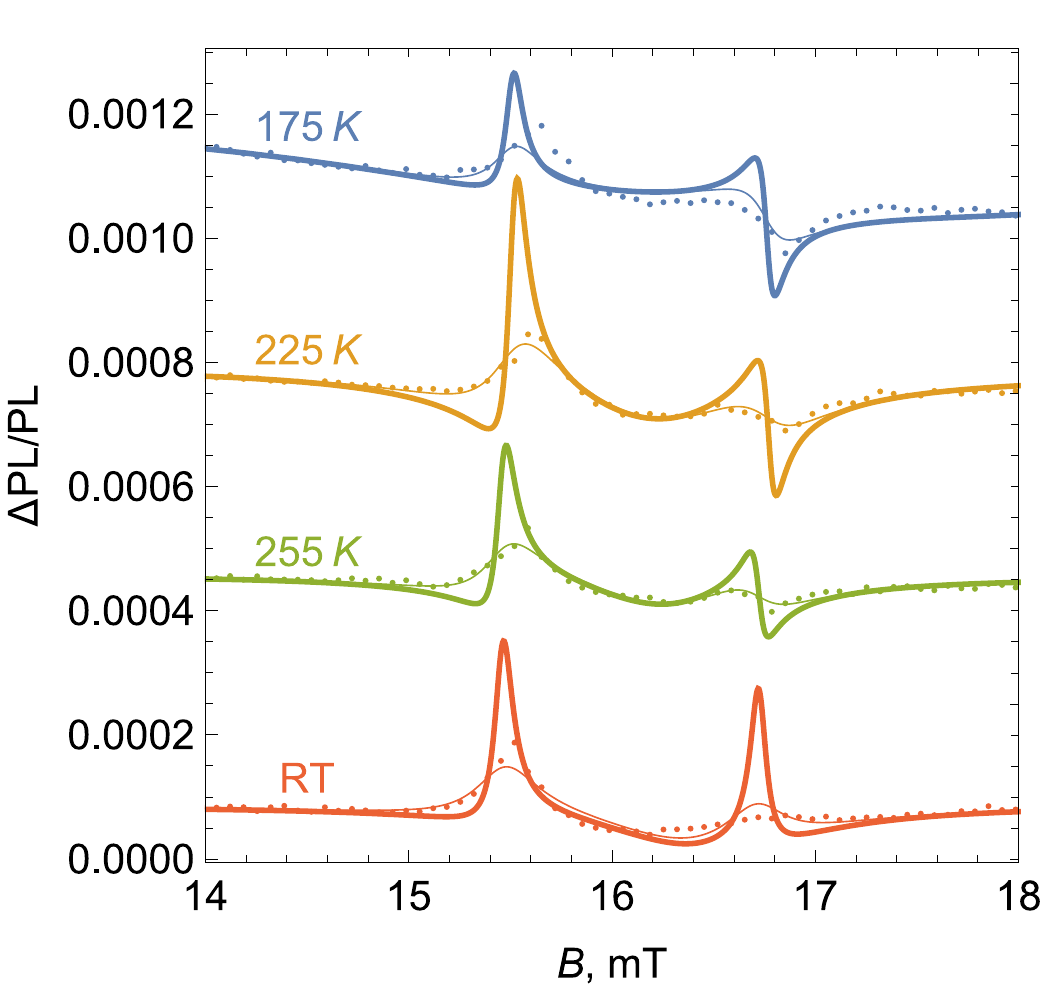}
\caption{Experimental ODMR spectra (points) and the result of their fitting with Eq.~\eqref{eq:dPL} (thin lines). Thick lines show the calculation made for a smaller value of the inhomogeneous broadening $W_g = 2.5\,$MHz, while all other parameters remain the same as in the fit.}
\label{fig:inh}
\end{figure}

To model the ODMR spectra, we used a sum of two signals, corresponding to the $-3/2 \leftrightarrow +1/2$ and $+3/2 \leftrightarrow -1/2$ transitions. At each temperature, we used $\Omega^{(g,e)}_z$ equal to the spin splitting of the corresponding states, and all other parameters were the same for both signals:
\begin{table}[h]
\begin{tabular}{ccccccc}
$T$, K & $\Omega^{(e)}_R/\Omega^{(g)}_R$, $10^2$ & $P$, MHz & $\Gamma$, MHz & $W_g$, MHz & $\gamma_g$, kHz &  $\gamma_e$, MHz \\
\hline
 175 & -4.6 & 0.38 & 86 & 7.7 & 0.04 & 0.4 \\
 225 & -4.9 & 0.65 & 260 & 10.7 & 0.12 & 1.2 \\
 255 & -3.5 & 0.76 & 240 & 9.3 & 0.21 & 2.1 \\
 300 & -1.2 & 1 & 250 & 10 & 0.46 & 4.6 \\
\end{tabular}
\end{table}

Here, $\gamma_g$ was taken from the temperature dependence of the spin relaxation rate measured in Ref.~\onlinecite{Simin:2017iw}. We assumed $\gamma_e = 10^4 \,\gamma_g$, and the other parameters were fitted to obtain the best match with the experimental data. The result of the fits are shown as thin curves in Fig.~\ref{fig:inh}. For comparison, we also show as thick curves the same ODMR spectra, but with a smaller inhomogeneous broadening $W_g = 2.5\,$MHz. In this case, the amplitude of the resonances increases and their asymmetry becomes more pronounced.


%